\documentclass[twocolumn,showpacs,pra,longbibliography]{revtex4-1}
\usepackage{graphicx}
\usepackage{latexsym}

\begin{document}

\title{Waves in intensity coherence of evolving intense twin beams}

\author{Radek Machulka}
\email{radek.machulka@upol.cz} \affiliation{Joint Laboratory of Optics,
Institute of Physics of the Czech Academy of Sciences, 17. listopadu 50a,
771~46 Olomouc, Czech Republic}

\author{Jan Pe\v{r}ina Jr.}
\email{jan.perina.jr@upol.cz} \affiliation{Joint Laboratory of Optics, Faculty of Science, Palack\'{y} University, 17.
listopadu 12, 771~46 Olomouc, Czech Republic}

\author{Ond\v rej Haderka}
\affiliation{Joint Laboratory of Optics, Faculty of Science, Palack\'{y} University, Czech Republic, 17. listopadu 12,
771~46 Olomouc, Czech Republic}

\author{Alessia Allevi}
\affiliation{Dipartimento di Scienza e Alta Tecnologia, Universit\`a degli Studi dell'Insubria, Via Valleggio 11, 22100
Como, Italy}

\author{Maria Bondani}
\affiliation{Istituto di Fotonica e Nanotecnologie, Consiglio Nazionale delle Ricerche, Via Valleggio 11, 22100 Como,
Italy}

\begin{abstract}
Strong correlations between the signal and idler beams imprinted during their
generation dominantly determine the properties of twin beams. They are also
responsible for the waves in intensity coherence observed in the wave-vector
space of a twin beam propagating in a nonlinear crystal in the regime with pump
depletion. These waves start to develop at certain twin-beam intensity and move
from the signal and idler beam centers towards their tails. They manifest
themselves via the change of coherence volume monitored in the far field by the
measurement of local modified $ \bar{g}^{(2)} $ function, which acts as a
sensitive and stable tool for investigating field intensity coherence.
\end{abstract}

\maketitle

\section{Introduction}

Coherence theory of light \cite{Born1980,Perina1985} established
in the 1950s completed the development of
classical optics and gave the physicists a tool for investigating
classical fields under the most general conditions. Importance of
the classical coherence theory of light valid in general for
intense fields, was also emphasized when its close similarity with
the quantum coherence theory of light \cite{Mandel1995,Perina1991},
needed for weak nonclassical fields, was recognized using the theory of
coherent states \cite{Glauber1963}. The general coherence theory
allows us to successfully describe all possible optical fields,
ranging from completely chaotic fields originating in spontaneous
emission to maximally coherent fields that arise in
stimulated emission. Wide possibilities for changing
coherence properties of optical fields are provided by nonlinear
(quantum) optics \cite{Mandel1995,Perina1991,Boyd2008} in the
framework of which mutual interactions of optical fields
are analyzed. Nonlinearity then opens the door to the
generation of nonclassical fields.

The generation of intense twin beams (TWBs), composed of a signal beam and an
idler beam with shared photon pairs, emerging in the optical parametric process
belongs to this area. Tight pairing of photons in TWBs even at the macroscopic
level has been experimentally confirmed in
Refs.~\cite{Jedrkiewicz2004,Bondani2007,Blanchet2008,Brida2009a} and used for
ghost \cite{Gatti2008} and quantum \cite{Brida2010a} imaging. The behavior of
TWBs under various conditions was addressed in
Refs.~\cite{Gatti2003,Brambilla2004,Brambilla2010,Caspani2010,Christ2011,Stobinska2012,Perez2014,Chekhova2015,Sharapova2015,Cavanna2016,Chekhova2018}
in the usual regime with strong (un-depleted) pump. In this regime, TWB
coherence increases as its intensity grows \cite{Gatti2003,Brambilla2004} due
to the prevailing stimulated emission. On the other hand, experimental
investigations of Ref.~\cite{Allevi2014a} revealed that, in the regime with
pump depletion, both spatial and spectral intensity coherence of a TWB do not
necessarily have to increase with the increasing TWB intensity
\cite{Allevi2014a}. Numerical statistical simulations of the solution of the
nonlinear Maxwell equations reported in \cite{Allevi2014a} confirmed that
pump-beam depletion leads to the fact that a TWB reaches its largest coherence
at certain intensity depending on the configuration of the nonlinear
interaction. The quantum theory of TWB generation
\cite{Law2000,Law2004,Fedorov2014}, extended into the regime with pump
depletion \cite{PerinaJr2016,PerinaJr2018b}, provided the physical explanation
by decomposing the nonlinear interaction into many independent spatio-spectral
modes \cite{PerinaJr2015} and analyzing their internal dynamics. Moreover, the
theory in Ref.~\cite{PerinaJr2016} pointed out at the existence of multiple
coherence maxima in different intensity regimes of TWBs \cite{PerinaJr2016a}
and coherent components in otherwise chaotic TWBs \cite{PerinaJr2016b}.
Further, detailed analysis of local coherence in a TWB, provided by a model
generalizing that of Ref.~\cite{PerinaJr2016}, revealed interesting behavior of
local intensity coherence across the TWB profile during its evolution
\cite{PerinaJr2019} whose observation is reported here. We note that the actual
TWB intensities at which such a behavior is observed strongly depend on the
value of the effective nonlinear coefficient, the interaction length as well as
the pump-beam parameters \cite{PerinaJr2016}. A pulse with spectral width
around 1~nm and transverse profile of several hundreds of $ \mu $m containing $
10^{14}-10^{15} $ photons allows to reach, in a BBO crystal several mm long,
the regime with pump depletion and to generate in the process of parametric
down-conversion a TWB containing $ 10^{12}-10^{14} $ photon pairs. On the other
hand, pump pulses by several orders in magnitude weaker suffice to reach the
regime of pump depletion when parametric down-conversion occurs in nonlinear
photonic structures especially in nonlinear waveguides. A large number of
independent spatio-spectral modes participating in parametric down-conversion
represents the second necessary requirement for the observation of such a
behavior.

A TWB, as it propagates along a nonlinear crystal, starts to develop from the
vacuum state due to spontaneously emitted photon pairs that take their energy
from a strong pump beam. At larger TWB intensities, stimulated emission of
photon pairs prevails and this results in gradual improvement of TWB coherence,
both in the spectral and (radial and azimuthal) wave-vector domains. We note
that the TWB coherence is characterized, in general, by a coherence volume
\cite{Perina1991} spanned in the above three domains and determined in all the
points of signal and idler beam profiles. Until the pump beam starts to be
depleted, it holds that the greater the local signal (or idler) beam intensity
the better the local coherence, as expected \cite{Gatti2003,Brambilla2004}.
However, when the pump beam is being depleted, dramatic changes in the
evolution of TWB coherence occur: The area with the best coherence (i.e. with
the largest coherence volume), originally localized in the signal (and idler)
beam center starts to move towards the beam tails as the overall TWB intensity
gradually increases. This forms waves in the coherence of the TWB that move in
the signal and idler frequency and radial wave-vector beam profiles, as the TWB
propagates along the crystal. These waves were recently theoretically predicted
in Ref.~\cite{PerinaJr2019}. We note that such waves in the coherence do not
propagate along the azimuthal wave-vector axis in a radially symmetric TWB
owing to constant signal- and idler-beam  intensities along this axis. Here, by
measuring the local modified $ \bar{g}^{(2)} $ function, we experimentally
investigate the behavior of waves in the TWB coherence.

\section{Theoretical model}

In the model developed in Ref.~\cite{PerinaJr2015a,PerinaJr2019}, the general
nonlinear momentum operator $ \hat{G}_{\rm int} $ of three mutually interacting
beams propagating along the $ z $ axis,
\begin{eqnarray}   
 \hat{G}_{\rm int}(z) &=& 2 \epsilon_0 \int dxdy \int_{-\infty}^{\infty} dt
   \nonumber \\
  & & \hspace{-13mm}
  \left[ \chi^{(2)} : \hat{E}^{(+)}_{\rm p}({\bf r},t) \hat{E}^{(-)}_{\rm
   s}({\bf r},t)
  \hat{E}^{(-)}_{\rm i}({\bf r},t) + {\rm H.c.} \right],
\label{1}
\end{eqnarray}
is approximately replaced by the following momentum operator $
\hat{G}_{\rm int}^{\rm av} $,
\begin{eqnarray}     
 \hat{G}_{\rm int}^{\rm av}(z) &=& i\hbar
  \sum_{m=-\infty}^{^\infty} \sum_{l,q=0}^{\infty}
    K_{mlq} \hat{a}_{{\rm p},mlq}(z) \hat{a}_{{\rm s},mlq}^{\dagger}(z)
   \hat{a}_{{\rm i},mlq}^{\dagger}(z) \nonumber \\
  & & \mbox{} + {\rm H.c.},
\label{2}
\end{eqnarray}
that effectively decomposes the three beams into many independent
triplets of modes constituted by modes from each interacting beam.
In Eq.~(\ref{1}), symbol $ \hat{E}^{(+)}_{\rm p} $ [$
\hat{E}^{(-)}_{\rm s} $, $ \hat{E}^{(-)}_{\rm i} $] means the
positive- [negative-] frequency part of the pump [signal, idler]
electric-field operator amplitude, $ \chi^{(2)} $ stands for the
second-order susceptibility tensor, $ \epsilon_0 $ is the
permittivity of vacuum, $ {\rm H.c.} $ replaces the Hermitian
conjugated term, and $ {\bf r} = (x,y,z) $. Annihilation [creation]
operators $ \hat{a}_{{\rm p},mlq} $ [$
\hat{a}_{{\rm s},mlq}^\dagger $, $ \hat{a}_{{\rm i},mlq}^\dagger
$] in Eq.~(\ref{2}) belong to independent spatio-spectral modes in
the corresponding beams indexed in azimuthal ($ m $) and radial ($
l $) wave-vector and frequency ($ q $) variables, $ \hbar $ stands
for the reduced Planck constant.

In the model, different effective nonlinear coupling constants $ K_{mlq} $ are
assigned to individual mode triplets composed of the modes with different
ability to build coherence. This gives us the key for the explanation of the
coherence behavior: The initial exponential increase of intensities in the
signal and idler modes of mode triplets leads to the gradual dominance of those
modes experiencing the largest coupling constants, and thus having the ability
to build the best coherence in the beam center. This is manifested by the
decrease in the number of TWB modes \cite{PerinaJr2016,Triginer2019}. However,
as already the classical solution of the three-mode interaction shows
\cite{PerinaJr2019}, the energy of pump modes in individual mode triplets is
gradually completely depleted in the nonlinear interaction and back-flow of the
energy towards the pump modes follows. This leads to the gradual loss of the
dominance of the signal and idler modes with the largest coupling constants.
This results in the enhancement of the role of the modes with smaller coupling
constants and weaker ability to build coherence in the beam center. As a
consequence, increase in the number of TWB modes occurs. On the other hand, due
to the mutual phase synchronization, these modes are endowed with the ability
to build highly coherent areas at the beam tails. It holds that, the smaller
the coupling constants of the mode triplets prevailing in the TWB, the more
distant the area with the largest coherence from the beam center. This is the
way how the internal dynamics of mode triplets forms the waves in the coherence
of evolving TWBs (for more details, see Ref.~\cite{PerinaJr2019}).

A simplified qualitative physical explanation can be provided by considering
the nonlinear interaction locally in the wave-vector and frequency domains. In
the beam center, the largest local intensities occur and so the increase of
coherence, followed by the gradual loss of coherence, develops the fastest. On
the other hand, local intensities at beam tails are smaller, so the largest
local coherence is reached later. The more distant the area from the beam
center, the lower the local intensities and so the later the largest local
coherence is reached.

\section{Experimental arrangement}

\begin{figure} 
 \centering
 \includegraphics[width=\linewidth]{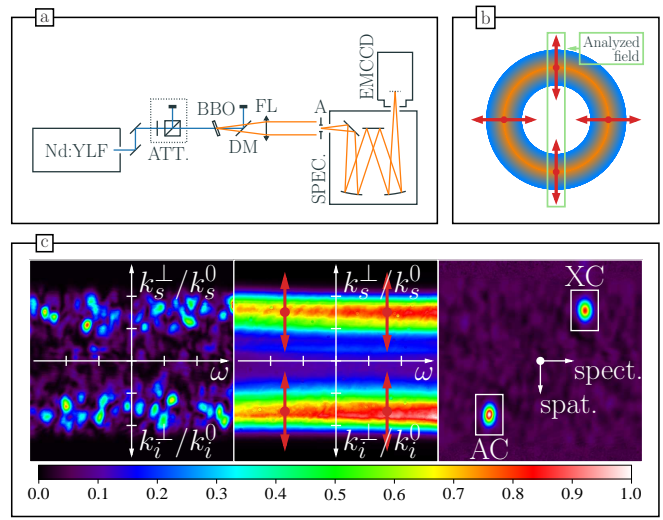}
 \caption{(a) Experimental setup used for the determination of spatio-spectral
  intensity correlations in a TWB. Nd:YLF: pump laser; ATT.: varying pump-beam attenuation; BBO: nonlinear crystal;
  DM: dichroic mirror to deflect the pump beam; FL: Fourier lens with 100-mm focal length; A: aperture with vertical rectangular slit;
  SPEC.: imaging spectrometer; EMCCD: electron-multiplying CCD camera.
  (b) Ring formed by the signal-beam intensity in the far field for central frequency.
  (c) Output of the imaging spectrometer. From left to right: single-shot image of both beams showing the speckle-like pattern with correlated
  grains, multiply-exposed image forming two horizontal strips (upper strip: signal beam, lower strip: idler
  beam) and scheme for the determination (observation) of intensity auto- (AC) and cross-
  (XC) correlations; color bar below quantifies the intensities in arbitrary
  units. In (c), the frequency axis $ \omega $ extends from 677.3~nm to 718.7~nm, the
  signal ($ k_{\rm s}^\perp $) and idler ($ k_{\rm i}^\perp $) radial wave-vector
  axes cover the range from 0~mrad to 40~mrad. In (b) and (c), red arrows
  indicate the propagation of waves in TWB coherence at increasing TWB intensity. They
  start to develop above certain threshold pump-power $ P_{\rm th} $.}
\label{fig1}
\end{figure}

The waves in TWB coherence, that develop in the radial-wave vector domain
during the TWB evolution along the nonlinear crystal towards the larger
intensities, were observed in an experimentally feasible arrangement in which a
nonlinear crystal of fixed length $ L $ is pumped by laser pulses with
gradually increasing power $ P $. The TWB coherence is analyzed beyond the
crystal as it varies with the increasing power $ P $. This approach relies on
the fact that the three-mode interaction in the considered configuration
evolves according to the parameter $ L\sqrt{P} $ \cite{Boyd2008,PerinaJr2019},
in which an increase of $ L $ can be equivalently substituted by an increase of
$ \sqrt{P} $. In detail, the experiment \cite{Allevi2014a} was performed in the
setup shown in Fig.~\ref{fig1}(a) using a type-I BBO crystal of length $ L=5
$~mm cut for frequency degenerate slightly non-collinear interaction. The
crystal was pumped by the third-harmonic pulses (349~nm, 4.5-ps pulse duration)
of a mode-locked regeneratively amplified Nd:YLF laser (High-Q-Laser) running
at 500~Hz. Intensity spectrum of the pump pulses was 1.2~nm wide [full width at
half maximum (FWHM)]. It was sufficiently broad to justify the application of
the model discussed above. A half-wave plate followed by a polarizing-cube beam
splitter was used to change the pump power $ P $. The corresponding parts of
the signal- and idler-beam emission cones [see Fig.~\ref{fig1}(b)] were
collected by a lens with 100-mm focal length and then collimated in the plane
of the vertical slit of an imaging spectrometer (Andor Shamrock SR-300i, 1200
lines/mm grating). The angularly dispersed far-field radiation was recorded,
shot by shot, by a synchronized EMCCD camera (iXon Ultra 897, Andor) operated
at full-frame resolution (512x512 pixels, 16x16~$\mu$m$^2$ pixel size). The
presence of intensity correlations between the signal and idler beams was
verified by observing symmetrically-positioned speckles around the degenerate
wavelength and slightly non-collinear emission ($ \theta_{\rm s} = 1 $~deg), as
shown in Fig.~\ref{fig1}(c). While the spectral correlations are observed in
the horizontal directions, the spatial radial (wave-vector) correlations are
visible in the vertical direction in Fig.~\ref{fig1}(c).

\section{Coherence and intensity auto-correlation function}

In principle, the adopted experimental arrangement allows us to directly follow
the evolution in TWB coherence only in two degrees of freedom, radial
wave-vectors and frequencies, via the measurement of the profiles of intensity
auto-correlation functions. However, also the changes of the local coherence
volume can be inferred from the measurement of the normalized intensity
auto-correlation function $ \bar{g}^{(2)} $ [see Eq.~(\ref{3}) below] performed
in a detection volume with a suitably chosen finite size [determined by the
area of (grouped) pixel(s) of the CCD chip and width of the slit in the beams'
transverse plane that defines the analyzed fields]. As the simple 1D model
presented below shows, the extension of the 3D detector in at least one
dimension has to be comparable to the corresponding coherence length to allow
for the observation of changes in this coherence length via the observation of
the $ \bar{g}^{(2)} $ function. Since the direct observation of coherence
changes in radial wave-vector and frequency domains via the corresponding
intensity auto-correlation functions $ A_{{\rm s},k} $ and $ A_{{\rm s},\omega}
$, respectively, requires an experimental geometry with 'tiny pixels' on the
camera to reach sufficient precision, we used a suitably wide slit in the
vertical plane of the input to the imaging spectrometer to allow for changes in
the $ \bar{g}^{(2)} $ function.

\begin{figure}         
 \centering
 \includegraphics[width=\linewidth]{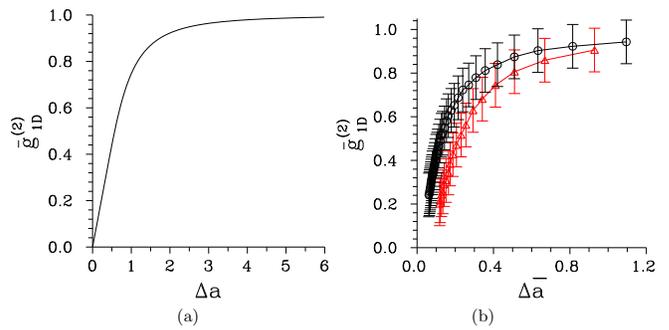}
 \caption{Functions (a) $ \bar{g}^{(2)}_{\rm 1D}(\Delta a) $ from Eq.~(\ref{4}) of the Gaussian model and
  (b) $ \bar{g}^{(2)}_{\rm 1D}(\Delta \tilde{a}) $ derived from the experimental data by
  grouping the pixels in the spectral (red $ \triangle $) and radial wave-vector (black $ \circ $)
  directions as they depend on normalized coherence lengths $ \Delta a = \Delta A/d $ and
  $ \Delta \tilde{a} = \sqrt{\ln(2)}\Delta a $. In (b), solid curves connect the experimental points.}
\label{fig2}
\end{figure}

The normalized modified $ \bar{g}^{(2)} $ function measured in a given
detection volume $ V $ is defined as
\begin{equation}  
 \bar{g}^{(2)} = \frac{ \langle (\Delta W)^2\rangle }{ \langle
 W\rangle^2} = \frac{ \langle W^2\rangle }{ \langle
 W\rangle^2}-1 = g^{(2)} -1
\label{3}
\end{equation}
in terms of the moments of the overall detection intensity $ W = \int_V d{\bf
r} I({\bf r}) $ and its fluctuation $ \Delta W = W - \langle W\rangle $ or
using the usual $ g^{(2)} $ function. We note that, according to the definition
in Eq.~(\ref{3}), $ \bar{g}^{(2)} =1$ for a single-mode thermal beam. A general
relationship between the $ \bar{g}^{(2)} $ function and local beam coherence
can be understood by considering an $ M $-mode thermal optical field with
equally populated modes defined in the detection volume. In this case, we have
$ \bar{g}^{(2)} = 1/M $ \cite{Perina1991}. Thus the larger the function $
\bar{g}^{(2)} $ the smaller the number $ M $ of local modes and consequently
the better the local coherence (the greater the coherence volume). To quantify
the relationship between the modified $ \bar{g}^{(2)} $ function and coherence
volume, we first consider a 1D model in which a pixel of extension $ 2d $ is
illuminated by a spatially stationary beam with (point) intensity $ I(x) \equiv
I_0 $ and intensity spatial correlation function $ \langle \Delta I(x) \Delta
I(x') \rangle \equiv {\cal I}(x-x') $, with $ \Delta I(x) = I(x) - I_0 $.
Assuming a Gaussian intensity auto-correlation function $ {\cal I}(x) =
\exp[-x^2/(\Delta A)^2 ] $ with $ \Delta A $ giving the coherence length, the
corresponding modified $ \bar{g}^{(2)}_{\rm 1D} $ function attains the form:
\begin{eqnarray} 
 \bar{g}^{(2)}_{\rm 1D} &=& \frac{ \int_{-d}^{d}dx \int_{-d}^{d} dx' \langle \Delta I(x) \Delta I(x') \rangle
   }{ \left[ \int_{-d}^{d}dx \langle I(x) \rangle \right]^2 } \nonumber \\
 &=& \frac{\sqrt{\pi}}{2} \frac{\Delta A}{d} {\rm erf}\left( \frac{d}{\Delta A} \right) ;
\label{4}
\end{eqnarray}
$ {\rm erf}(x) = (2/\sqrt{\pi}) \int_{0}^{x} dt \exp(-t^2) $. According to
Eq.~(\ref{4}), we have $ \bar{g}^{(2)}_{\rm 1D} \rightarrow 1 $ for $ \Delta A
\gg d $, i.e. a fully spatially coherent field is composed of just one mode. On
the other hand, $ \bar{g}^{(2)}_{\rm 1D} \rightarrow 0 $ in the opposite case $
\Delta A \ll d $ of a weakly spatially coherent field composed of many modes.
The graph of $ \bar{g}^{(2)}_{\rm 1D} $ function of Eq.~(\ref{4}), plotted in
Fig.~\ref{fig2}(a), points out a strong sensitivity of the $ \bar{g}^{(2)}_{\rm
1D} $ function on the normalized coherence length $ \Delta a \equiv \Delta A/d
$ for $ \Delta a_x \in \langle 0.2,3 \rangle $. Assuming for simplicity the
factorized intensity correlation functions in 3D model, the above 1D model is
appropriate in all the three dimensions: radial and azimuthal wave-vectors and
spectral. For comparison, the behavior of $ \bar{g}^{(2)}_{\rm 1D} $ function
derived from the measurement in frequency and radial wave-vector directions by
fixing the coherence length (signal-beam center, at threshold power $ P_{\rm
th} $) and varying the detector width (grouping of pixels) is shown in
Fig.~\ref{fig2}(b).

\section{Observation of waves in TWB coherence}

\begin{figure}         
 \centering
 \includegraphics[width=\linewidth]{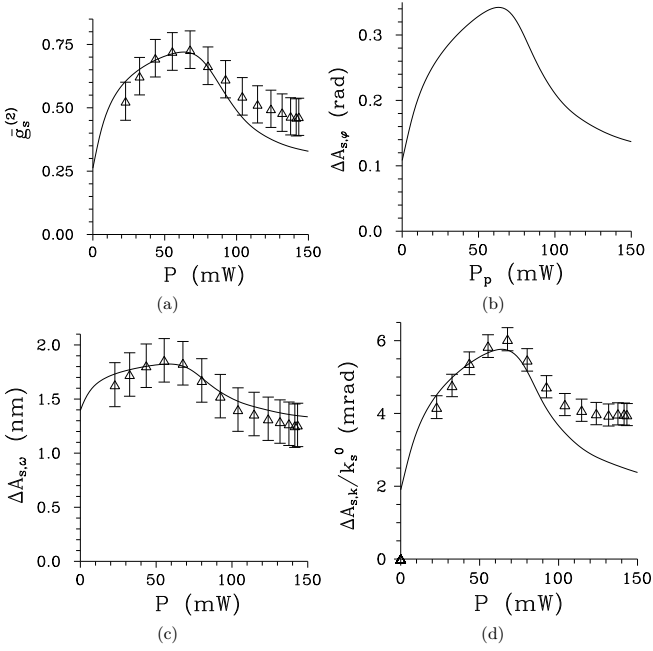}
 \caption{(a) Function $ \bar{g}^{(2)} $ measured at individual pixels and
  widths (b) $ \Delta A_{{\rm s},\varphi} $,
  (c) $ \Delta A_{{\rm s},\omega} $ and (d) $ \Delta A_{{\rm s},k} $
  of, in turn, azimuthal wave-vector, spectral and radial wave-vector intensity auto-correlation functions
  (FWHM) for the signal beam at central frequency $ \omega_{\rm s}^0 $ and
  central radial wave vector $ k_{\rm s}^{0} $ as they depend on pump power
  $ P $. Experimental data are plotted as isolated symbols with
  error bars, solid curves originate in the theoretical model \cite{PerinaJr2019}.}
\label{fig3}
\end{figure}

To document close similarity in the behavior of the modified $ \bar{g}^{(2)} $
function and widths $ \Delta A_{{\rm s},\varphi} $, $ \Delta A_{{\rm s},\omega}
$ and $ \Delta A_{{\rm s},k} $ of intensity auto-correlation functions in the
azimuthal wave-vector ($ \varphi $), frequency ($ \omega $) and radial
wave-vector ($ k $) directions, respectively, we draw their dependence on pump
power $ P $ in the signal-beam center in Fig.~\ref{fig3}. All four quantities
increase at increasing pump power $ P $ below the threshold power $ P_{\rm th}
\approx 70 $~mW ($ 2.45 \times 10^{14} $ photons per single pump pulse) as a
consequence of the gradual increase of TWB coherence in all three directions.
Then, for higher pump powers, coherence is partially lost due to the gradual
increase of the number of effectively-populated TWB modes, as discussed above.
The theoretical curves displayed in Fig.~\ref{fig3} originate in the model of a
spatially and spectrally multi-mode TWB of Ref.~\cite{PerinaJr2019}, in which
we assume a Gaussian pump pulsed beam at wavelength 349~nm, with repetition
rate 500~Hz, being 230~$ \mu $m wide (intensity FWHM, at the input crystal
plane) and having temporal spectrum 1.2~nm wide (intensity FWHM) together with
a BBO crystal 4~mm long. The crystal phase matching was such that the signal
beam leaved the crystal at $ \theta_{\rm s}^0 = 1.0 $~deg. The model predicts $
2.29 \times 10^{12} $ photon pairs per single pump pulse at $ P = 70 $~mW.

\begin{figure}         
 \centering
 \includegraphics[width=\linewidth]{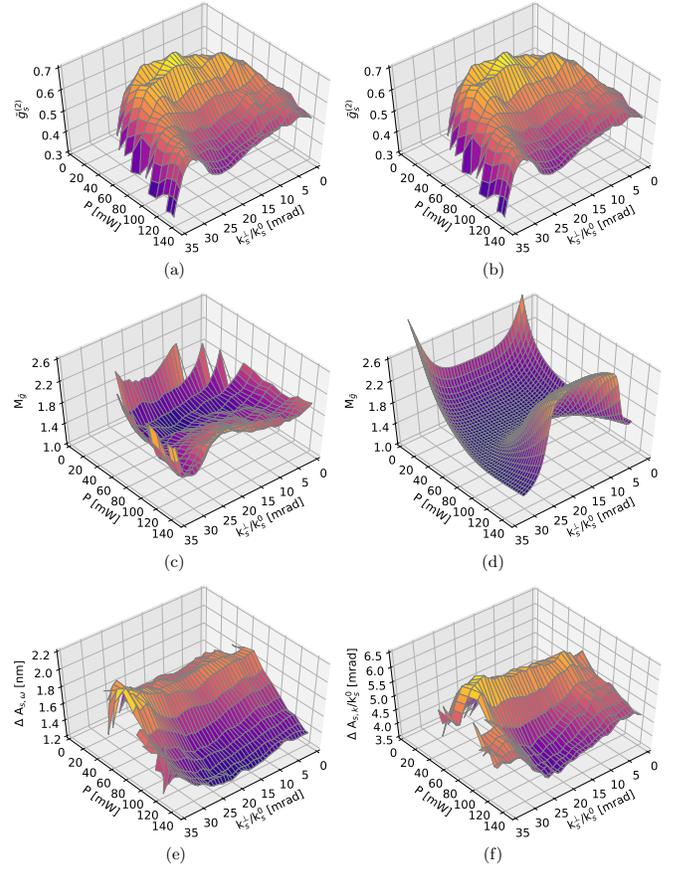}
 \caption{(a) Experimental and (b) theoretical $ \bar{g}^{(2)} $
  function for individual pixels, (c) experimental and (d) theoretical number
  $ M_g = 1/\bar{g}^{(2)} $ of local modes and experimental widths (e) $ \Delta A_{{\rm s},\omega} $ and
  (f) $ \Delta A_{{\rm s},k} $ of frequency and radial wave-vector, respectively, intensity autocorrelation
  functions [FWHM] as they depend on normalized radial wave vector $ k_{\rm s}^\perp/k_{\rm s}^0 $
  and pump power $ P $ for the central signal-beam frequency $ \omega^0_{\rm s} $.
  Relative errors are better than 5\% in (f), 11\% in (e) and 12\% in (a,c).}
\label{fig4}
\end{figure}

\begin{figure}         
 \centering
 \includegraphics[width=0.7\linewidth]{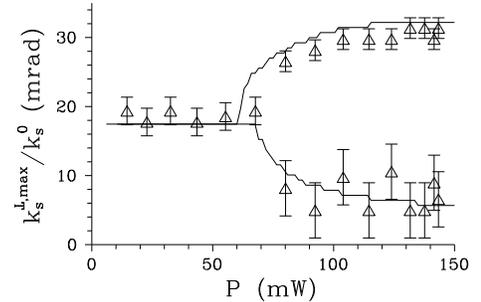}
 \caption{Positions of maxima of $ \bar{g}^{(2)} $ function drawn in
  Figs.~\ref{fig4}(a) and (b) in signal-beam radial wave vector
  $ k_{\rm s}^\perp $ as they depend on pump power $ P $. Experimental
  (theoretical) data are plotted as isolated symbols (by solid curves).}
\label{fig5}
\end{figure}

\begin{figure}        
 \centering
 \includegraphics[width=\linewidth]{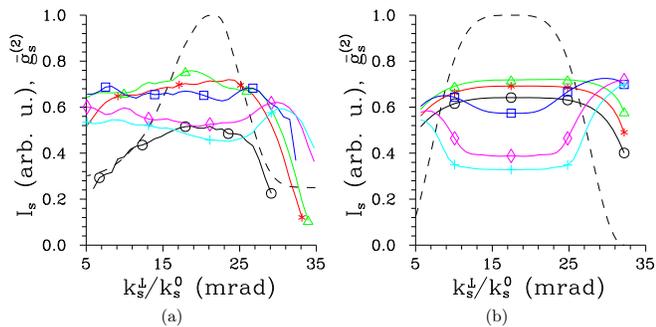}
 \caption{(a) Experimental and (b) theoretical profiles of the signal-beam $ \bar{g}^{(2)} $ function
   drawn in Fig.~\ref{fig4} [relative errors better than 12\%] along normalized radial wave vector
   $ k_{\rm s}^\perp/k_{\rm s}^0 $ for powers $ P $ equal to 23~mW [30~mW] (solid curves with black $ \circ $),
   43~mW [45~mW] (red $ \ast $), 55~mW [60~mW] (green $ \triangle $), 80~mW [90~mW] (blue $ \Box $),
   115~mW [120~mW] (magenta $ \diamond $) and 142~mW (150~mW) (cyan +) drawn in (a) [(b)].
   Signal-beam intensity profile $ I_{{\rm s}, k} $ [relative error 1\%] is drawn by dashed curves.}
\label{fig6}
\end{figure}

\begin{figure}         
 \centering
 \includegraphics[width=\linewidth]{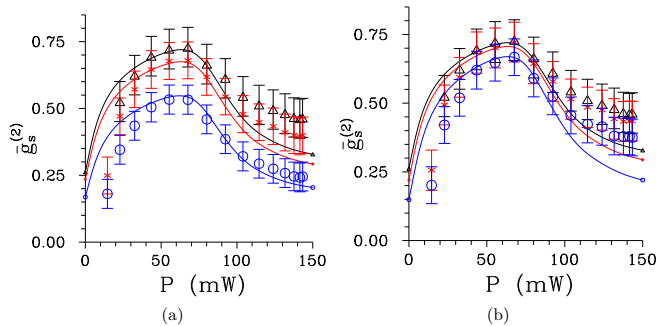}
 \caption{Functions $ \bar{g}^{(2)} $ obtained by grouping 1 (black $ \triangle $), 4 (red $ \ast $),
  and 8 (blue $ \circ $) pixels in (a) frequency and (b) radial wave-vector directions
  for the signal beam at central frequency $ \omega_{\rm s}^0 $ and
  central radial wave vector $ k_{\rm s}^{\perp 0} $ as they depend on pump power
  $ P $. Experimental data are plotted as isolated symbols with
  error bars, solid curves originate in the theoretical model.}
\label{fig7}
\end{figure}

The experimental modified $ \bar{g}^{(2)} $ function and the corresponding
local number $ M_g \equiv 1/\bar{g}^{(2)} $ of modes per single pixel measured
in the signal-beam strip at the central frequency $ \omega^0_{\rm s} $ are
drawn in Figs.~\ref{fig4}(a) and (c) as a function of the varying normalized
radial wave vector $ k_{\rm s}^\perp/k_{\rm s}^0 $ and pump power $ P $.
Their theoretical counterparts are also shown in Figs. \ref{fig4}(b) and (d),
respectively. Splitting of the maximum of the $ \bar{g}^{(2)} $ function in the
radial wave-vector direction ($ k_{\rm s}^\perp/k_{\rm s}^0 $) for the pump
powers higher than $ P_{\rm th} \approx 70 $~mW is evident in the graph in
Fig.~\ref{fig4}(a). The maximum splits at $ P_{\rm th} $ into two local maxima
that move towards the low and high radial wave-vectors $ k_{\rm s}^\perp $,
i.e. towards the signal-beam tails. This movement at increasing pump power $ P
$, and hence at increasing overall TWB intensity, exhibits the propagation of
waves in TWB coherence during the TWB evolution. The positions of the (local)
maxima in the signal-beam radial wave-vector direction as they evolve at
increasing pump power $ P $ are drawn in Fig.~\ref{fig5}.
Similarly, typical profiles of $ \bar{g}^{(2)} $ function in the signal-beam
radial wave-vector direction are shown in Fig.~\ref{fig6} for fixed pump powers
$ P $.
The modulation of $ \bar{g}^{(2)} $ function along the signal radial
wave-vector direction is dominantly caused by the variation of coherence length
with the pump power in the azimuthal wave-vector direction. This is so since
the pixel widths in the frequency (0.083~nm) and radial wave-vector (0.16~mrad)
directions are by one order in magnitude smaller than the corresponding
coherence lengths. To verify the method of monitoring the TWB coherence via the
measurement of $ \bar{g}^{(2)} $ function, we summed the intensities from 4 and
8 neighboring pixels for both directions on the EMCCD and observed a decrease
in the values of $ \bar{g}^{(2)} $ function, as documented in Fig.~\ref{fig7}.
It holds that the wider the detection area on EMCCD, the smaller the $
\bar{g}^{(2)} $ function, the larger the number of local modes and so the worse
the local coherence in the detection area.

In the experiment, the waves in TWB coherence along the signal radial
wave-vector profile were also observed by directly monitoring the widths $
\Delta A_{{\rm s},\omega} $ and $ \Delta A_{{\rm s},k} $ of frequency and
radial wave-vector intensity autocorrelation functions. However, as evidenced
in Figs.~\ref{fig4}(e) and (f), these widths are determined from the
experimental data with considerably larger fluctuations than the $
\bar{g}^{(2)} $ function in Fig.~\ref{fig4}(a). Good 'stability' of the
measurement of $ \bar{g}^{(2)} $ function comes from its natural 3D character
(intensity in its definition is integrated over the detection volume) that
remains also in our configuration with the dominant influence from only the
azimuthal wave-vector direction. This, together with strong sensitivity to
coherence changes (see Fig.~\ref{fig2}) makes the measurement of $
\bar{g}^{(2)} $ function superior for the observation of coherence variations.

Before we conclude, we note that the waves in TWB coherence propagate also
along the spectral (frequency) profiles of the signal and idler beams,
similarly to those observed along the radial wave-vector profiles.
Nevertheless, the signal- and idler-beam spectra generated from the BBO crystal
are relatively wide and so only a fraction of them is experimentally analyzed
[see Fig.~\ref{fig1}(c)]. Such configuration is not suitable for the
observation of coherence waves. We also note that the agreement between the
experimental and theoretical behaviors of the analyzed $ \bar{g}^{(2)} $
functions is in some cases only qualitative [Figs.~\ref{fig4}(a,b), 6(a,b)].
The reason is that the theoretical model of Ref.~\cite{PerinaJr2019} does not
include birefringence of the BBO crystal (walk-off effects modelled by the
effective shortening of the interaction length) and considers pump beams with
Gaussian spatial and temporal spectra (contrary to multi-mode high-intense
experimental pump beams).

\section{Conclusions}

The waves in twin-beam coherence, developing as the twin beam evolves
(increases its energy) in the nonlinear process, were indirectly observed in
the signal-beam profile in the radial wave-vector direction using the
configuration with a crystal of fixed length illuminated by a pump beam with
varying power. In the far field of the signal and idler beams, i.e. in their
transverse wave-vector spaces, these waves attain a ring-shaped form. They
propagate from the beams' centers towards their tails in the radial wave-vector
direction as the twin-beam intensity increases during its evolution. Similar
waves occur also in the spaces spanned by azimuthal wave vector and frequency.
These waves in 2D spaces, in fact, represent projections of the waves in
twin-beam coherence formed in the signal- and idler-beam 3D wave-vector spaces.
The waves in twin-beam coherence were indirectly monitored via the suggested
measurement of changes of the local modified $ \bar{g}^{(2)} $ function
registered in the volume of individual pixels at EMCCD camera. Higher
sensitivity of the modified $ \bar{g}^{(2)} $ function to the changes of
coherence compared to direct measurement of widths of intensity
auto-correlation functions in three orthogonal directions was successfully
utilized.

\acknowledgements The authors thank the bilateral Czech-Italian project CNR-16-05 between CAS and CNR and projects
No.~18-08874S by GA \v{C}R (R.M. and O.H.) and No.~18-22102S by GA \v{C}R (J.P.). Support from the project
No.~CZ.1.05/2.1.00/19.0377 of M\v{S}MT \v{C}R is also acknowledged.

%

\end{document}